# Morphological Control of Linear Particle Deposits from the Drying of Inkjet-Printed Rivulets

N. Bridonneau,[1,2,3] G. Mattana,[2*] V. Noël,[2] S. Zrig[2*] and F. Carn[1*]

[1] Université de Paris, Laboratoire Matière et Systèmes Complexes, CNRS, UMR 7057, Paris, France.
[2] Université de Paris, Laboratoire Interfaces Traitements Organisation et Dynamique des Systèmes, CNRS, UMR 7086, 15 rue J-A de Baïf, F-75013 Paris, France
[3] Actual address: Université Paris-Saclay, Institut de Chimie Moléculaire et des Matériaux d'Orsay, CNRS, UMR 8182,  91405, Orsay, France.

[*]E-mail : florent.carn@univ-paris-diderot.fr, samia.zrig@univ-paris-diderot.fr, giorgio.mattana@univ-paris-diderot.fr

**ABSTRACT**

We studied the morphology of linear particle deposits obtained by inkjet-printing of silica nanoparticle suspension in drying condition where contact line depinning occurs. We show that this evaporation mode can be obtained by adjusting the particle concentration in different solvents. For isolated drops, deposited manually or by inkjet printing, drying induces the formation of two concentric rings in which particles self-assemble into a monolayer. For fused drops, our main result is that stable rivulets could be formed by drop overlap leading, after drying, to the formation of three parallel lines composed of a self-assembled particle monolayer. The three lines are of homogeneous thickness with two very thin outer lines (~ 1 μm width) and a wider central line (~ 20 μm width). We reveal how the width of the resulting lines is influenced by drop spacing in a predictable manner for a large experimental window knowing the drop size.

# 1. TABLE OF CONTENTS IMAGE

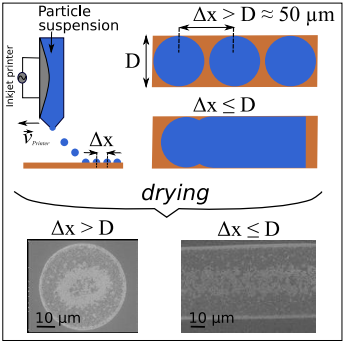

The drying of a drop of particle dispersion in open air on a solid substrate leads to a solid deposit which can take different forms such as coffee ring, stick-slip pattern, dot-like pattern or even uniform pattern.[1] This evaporative process is at the basis of a large number of coating applications.[2] Although uniform deposits have been mostly targeted so far, the formation of non-uniform deposits has nevertheless also been aimed in recent years. In the biomedical field, several studies tried to establish a relationship between certain pathological conditions and the shape of the deposit obtained by drying a biofluid.[3,4] Furthermore, the formation of coffee rings composed of metal nanoparticles in compact packing enabled the formation of Surface-Enhanced Raman Scattering sensors[5,6] or of conductive micrometric domains without post-annealing step.[7] The latter result has been exploited in the field of inkjet printing to quickly produce conductive lines of micrometric width with commercial ink-jet printer using nozzle with a diameter of 75 µm.[8–11] The method consists in depositing individual drops of nanoparticle dispersion on a surface with a droplet spacing allowing their coalescence to form stable liquid rivulets[12,13] that become a pair of solid "twin lines" after solvent evaporation and particle self-assembly at the pinned contact line.[14] This method is a simple alternative for forming nanostructures with high spatial resolution without the need for nozzle design or the application of an external electric field between the nozzle and substrate.[15] It has been shown experimentally that the width of a single line (w) can be minimized, down to 2 µm, by minimizing the drop size (D) knowing that $w \propto D$ for individual drops.[16] Qualitatively, this relation reflects that the solute initially dispersed in the drop, of volume proportional to $D^3$, is finally deposited in the external ring whose volume is proportional to $Dw^2$ so that $w \propto D$. Given the limited range of individual drop size (i.e. 20-80 µm), w can be further minimized by increasing the drop spacing to the limit of 'stable' coalescence,[13] increasing the temperature to speed up evaporation[17] and/or increasing the advancing contact angle.[18]
Surprisingly, past studies have given little consideration to the effect of particle concentration and have been only focused on the formation of twin lines resulting from an evaporation process where the contact line is always pinned. The purpose of the present work is to show that the morphology and dimensional characteristics of linear solid deposits can be modified by changing the evaporation mode. Our main result is that, 3 parallel and continuous lines, composed of particles in compact packing, with $w_{min} \approx 1.4$ µm and an inter-line spacing, d, down to 10 µm, could be obtained by choosing experimental conditions that allow contact line depinning during drying to form a central deposit between the external twin lines classically obtained by a "coffee ring" approach (Figure 1).

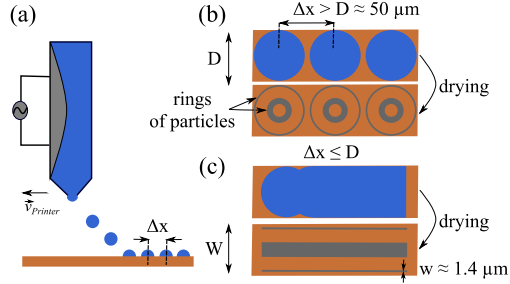

**Figure 1.** Naïve scheme of the inkjet printing process with (a) a cross-sectional view and (b) a top view for drop spacing (Δx) higher than sessile drop diameter (D). (c) Top view before and after drying for Δx ≤ D.

As a case study, we considered cationic silica nanoparticles, noted SiNP, (Klebosol® from Merck) and a smooth gold substrate coated by a self-assembled monolayer of 11-mercaptoundecanoic acid ($HS(CH_2)_{10}CO_2H$). The ink was prepared by dispersing SiNP in a solvent mixture composed of milliQ water, ethanol (99.9 %, Sigma-Aldrich), n-propanol (99.9 %, Sigma-Aldrich), and ethylene glycol (99.9 %, Sigma-Aldrich). The proportion of the different solvents was fixed in our experiment mixture as follows: water (48 vol.%), ethyleneglycol (32 vol.%), n-propanol (10 vol.%) and ethanol (10 vol.%). We choose this composition to improve the printing quality with our commercial inkjet printer based on the results of Kim et al.[9] The main characteristics of the different components of the system are given in the table 1. We used a Dimatix DMP-2800 printer with platen and cartridge temperature fixed at 30 °C. 10 pL cartridges and 20 µm nozzles were used and printing was carried out using a drop-spacing of 20 to 40 µm and a drop velocity of 7 m/s. Additional information concerning materials and methods are given in supplementary material.

| SiNP | R nm | ζ mV | Ink | $\Phi_{NP}$ vol.% | η mPa.s | γ mN.m$^{-1}$ | AuSub | Rms nm | Θ ° |
|---|---|---|---|---|---|---|---|---|---|
| | 37.5 | 50 | | 0.4 | 3.4 | 35 | | 2.8 | 60 |

**Table 1.** Main characteristics of the nanoparticles, ink and substrate used in this study with their acronym: particle radius measured by SEM-FEG (R), particle zeta potential (ζ) measured by Laser Doppler velocimetry, particle volumic fraction ($\Phi_{NP}$), ink dynamic viscosity at a shear rate of 2.5 s$^{-1}$, as measured with a cone (D = 4.8 cm) and plate geometry, surface tension (γ) of the inks measured with a Aqua Pi Kibron needle tensiometer, RMS roughness measured by AFM (Rms) and static contact angle measured with a Krüss DSA100 apparatus. All measurements were done at ambient condition (T = 25 °C and relative humidity ∼ 40 %).

We first observed dry patterns obtained by drying simple aqueous SiNP dispersion on the gold substrates in open air for particle initial concentrations varying between 0.01 and 100 g/L (i.e. $4.10^{-4} \leq \Phi_{vol.}\% \leq 4$). In this preliminary study, we deposited 0.5 μL using a manual micropipette in the cleanroom (25 °C, 40 % of relative humidity). Figure 2 shows typical images of the dry patterns obtained with a scanning electronic microscope equipped with a field emission gun (SEM-FEG).

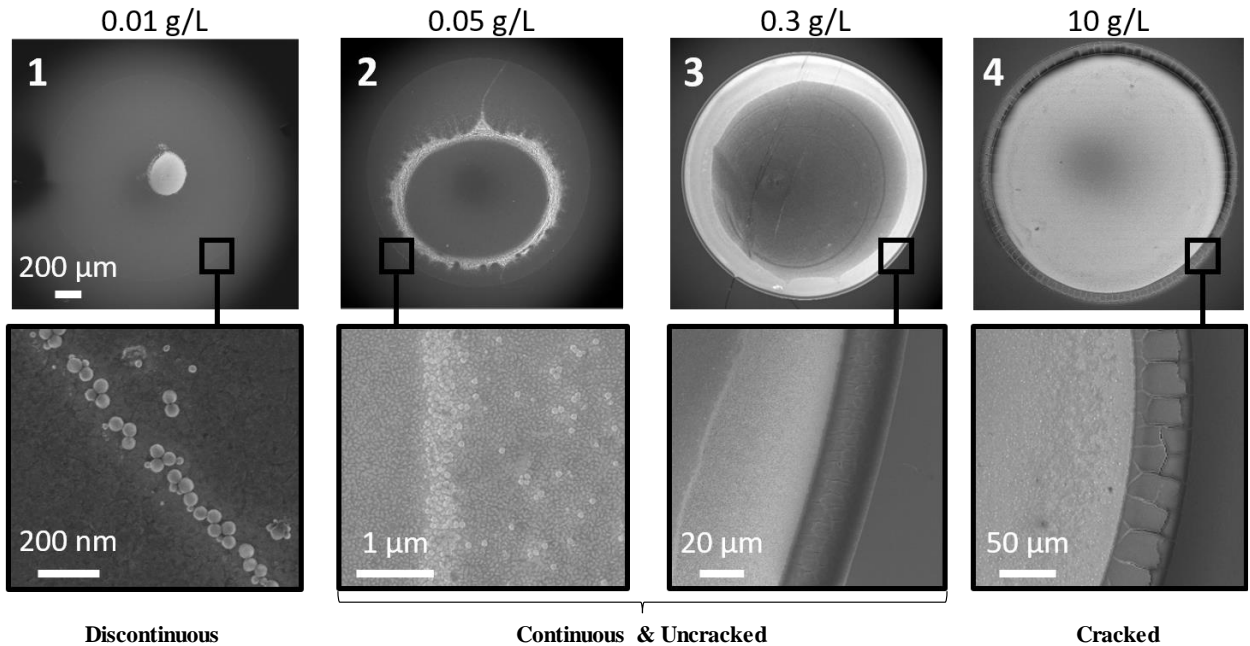

**Figure 2.** Typical SEM-FEG images at different magnifications of the dry patterns formed after the evaporation of sessile drops manually deposited (V = 0.5 μL) of SiNP suspensions at different initial concentrations as indicated. The scale bar is the same for all images in the top line. The images in the bottom line are enlargements of the image just above in the area of the external ring.

We have reproducibly observed a concentration-controlled shape transition sequence with different batches of particles and substrates. At low concentration, particles deposited preferentially in the center of the initial imprint of the drop, forming dot-like patterns. When particle concentration increased, particles started depositing at the drop periphery, forming a thick ring pattern, and finally a deposition over the entire surface was observed at high concentration.

In the dilute domain (i.e. [SiNP] $\leq 0.04 \pm 0.01$ g/L), noted 1 in figure 2, the dot-like pattern is composed of a continuous domain of particles compacted in a monolayer with holes reminiscent of "spinodal dewetting" surface patterns. A transition of pattern morphology toward 'thick' ring patterns occurred when the particle concentration is in the 0.05-0.10 g/L range (i.e. $\Phi_{vol.} \approx 2.10^{-3}$- $4.10^{-3}$ %). The patterns observed in this domain 2 present a peculiar structure taking the form of a thick ring whose outer diameter is much smaller than the diameter of the initial deposited droplet. The disk situated at the center of the thick ring is free of particles, whereas fine satellite rings could be detected beyond the outer edge. In these two first domains of concentration, the presence of concentric rings reflect a stick-clip evaporation process.[1,19] When the concentration is further increased we enter in the domain 3 where the ring widens until it reaches the initial edge of the initially placed drop for [SiNP] ~ 0.3 g/L (i.e. $\Phi_{vol.} \approx 1.2 \times 10^{-2}$ %). Beyond this concentration, the central disc, which until then had remained empty of particles is progressively covered with a monolayer of particles. The surface is completely covered from [SiNP] ~ 10 g/L (i.e. $\Phi_{vol.} \approx 0.4$ %). A ring of particles with an outer diameter corresponding to the initial diameter of the deposited drop was observed regardless of the particle concentration. This external ring is always separated from the rest of the deposit. As expected from previous studies, the width (w) of this external ring increases according to a power law with the particle concentration going from a fine deposit of a few particles wide (domain 1), to a compact multi-layer assembly (domains 2 & 3) and finally a compact, but cracked, assembly at high concentration (domain 4).

Most of previous 'inkjet' experiments on the formation of twinned lines by the so called "evaporation-driven convective particle self-assembly" were done in a concentration domain similar to the domain 3 where particles are packed in a continuous external ring without cracks. To the best of our knowledge, the domain 2 has never been the subject of studies in the inkjet printing field. However, the particle assembly observed in this domain meets several important criteria from an application point of view (i.e. continuous domain, particles in compact packing & no cracks) with a minimum quantity of particles, which is interesting considering the price of inks. Obviously, drying a 0.5 µL drop of water is different from drying a 10 pL drop of solvent mixture sprayed at ~ 7 m.s$^{-1}$ by a moving print head. Nevertheless, we were able to obtain inkjet deposits with a morphology similar to that obtained by manual deposition in the domain of concentration n°2. To do this we had to print with a drop-to-drop spacing ($\Delta x$) higher than the drop size (Figure 1b) and at a much higher particles concentration, around 10 g/L.

As shown in figure 3a, nice circular patterns were obtained repeatedly with diameters of 47 µm from measurements on SEM-FEG images. These patterns are composed of an external ring of width 1.2 µm and a central ring, which is not always circular, of average diameter 25 µm. Zooming in on the patterns, we observed that the particles in the central ring form a compact monolayer assembly with holes as in the previous case.

The external ring is also composed of a compact monolayer assembly without holes. In contrast to manually deposited droplets, the central ring obtained by inkjet printing is surrounded, inside and outside, by small clusters of nanoparticles forming monolayer islands.

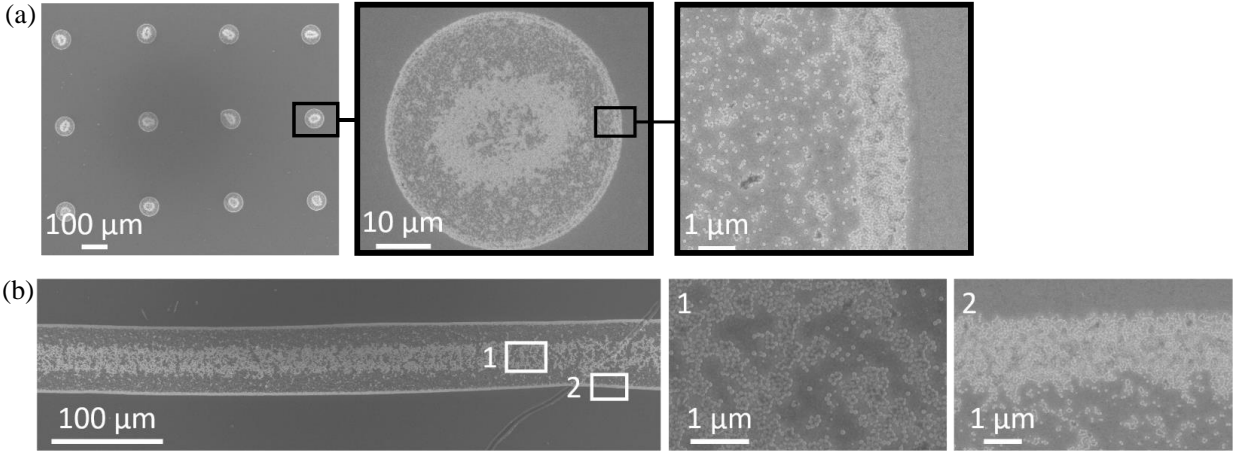

**Figure 3.** (a) Typical SEM images of dry deposits from unfused droplets deposited by inkjet printing, with $\Delta x > D = 47$ µm, at different magnifications. (b) Typical SEM images of dry deposits from fused droplets deposited by inkjet printing, with $\Delta x = 40$ µm $< D$, at different magnifications.

We then varied the drop-to-drop spacing between consecutive droplets by adjusting the drop frequency, f, and the translation speed, v (Figure 1a), as $\Delta x = v/f$, to form stable rivulets by drop overlap (i.e. $\Delta x \lesssim D$ with D, the drop size as shown in figure 1c). In this way we were able to form rivulets of width and length corresponding respectively to D and $n \times D$ with n the number of fused drops. After evaporation, a solid deposit of almost the same size as the rivulet is recovered (Figure 3b). This solid deposit consists of three continuous lines of uniform widths with two very thin twin lines defining the outer edges of the deposit and a thick central line. Locally, the particle assembly within the external lines is similar to that described above for the unfused drops with a width of the order of 1-2 µm. The central deposit has a diameter corresponding to the diameter of the central ring previously described for unfused drops. The local particle assembly is also similar but with a homogeneous particle density. Indeed, the only difference with the local structure observed for non-fused deposits is the disappearance of the diluted zone initially observed in the center of the rings. Small clusters of nanoparticles could be detected in between de thick central lines and the fine external lines. We point out that clusters of similar structures had been previously observed in the case of simple twin line deposits.[20,10] In addition, we have verified that working at a higher particle concentration, corresponding to domain 3 in figure 2, results in a drying condition where the contact line remains pinned in the same place until drying is complete. Under these conditions, we recover the solid deposits already reported in the literature, namely (i) the classic "coffee ring" deposit when $\Delta x > D$ and (ii) a pair of solid "twin lines" when $\Delta x \lesssim D$, as shown in figure S3.

Finally, we studied the dependence of the characteristic dimensions of the different lines with the drop-to-drop spacing. Regardless the drop-to-drop spacing we got three continuous lines of uniform widths with two very thin twin lines and a thick central line which shows the robustness of the three-lines deposition technique (Figure 4a).

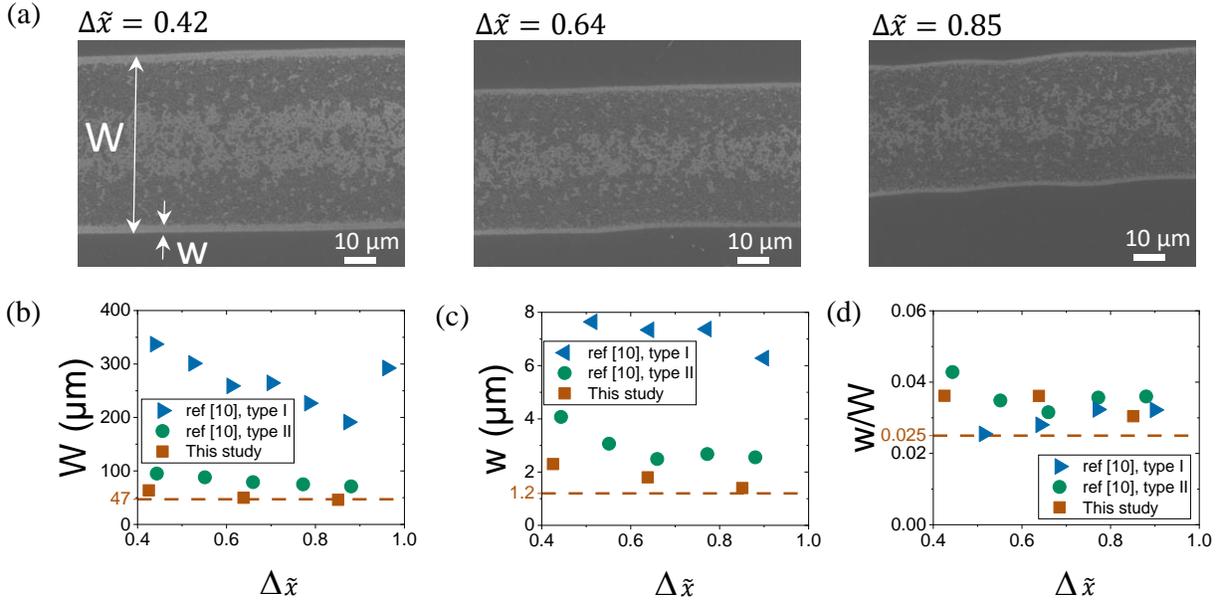

**Figure 4.** (a) Typical SEM-FEG images of dry deposits at different normalized drop spacing ($\Delta \tilde{x}$). (b) & (c) Evolution of the width of the full deposit (W) and of the external lines (w) with $\Delta \tilde{x}$ respectively. (d) Evolution of the ratio w/W with $\Delta \tilde{x}$. The dashed lines plotted in (b), (c) & (d) indicate the values of W, w and w/W for unfused droplets (see Fig. 2a) respectively. The study [10] was conducted with a dilute (1% wt.) aqueous dispersion of Ag particles (D = 77 nm) onto two types of glass substrates showing different advancing contact angles (i.e. $\theta_A^{type\ I} \approx 10\ °$ and $\theta_A^{type\ II} \approx 45\ °$).

To compare with previous studies, we normalized the drop spacing as $\Delta \tilde{x} = \Delta x / D$, where D is the drop size which is almost equal to the width of the solid deposit (W). As shown in figures 4a and 4b, the width of the whole deposit, W, decreases with $\Delta \tilde{x}$. The width of the external lines (w) follows the same trend toward the width of the unfused droplet external ring (Figure 4c). This agrees with previous results on twin lines although the range of $\Delta \tilde{x}$ under study is too narrow to verify the theoretically expected power law.[10]

Interestingly, the ratio between w and the width of the whole deposit (see Figure 4a), W, does not vary with $\Delta \tilde{x}$ as well as with the deposition condition (Figure 4d). Indeed, $w/W \approx 0.033$ for this study and for a previous study[10] conducted with a dilute (1% wt.) aqueous dispersion of Ag nanoparticles (D = 77 nm) onto two types of glass substrates showing different advancing contact angles (i.e. $\theta_A^{type\ I} \approx 10°$ and $\theta_A^{type\ II} \approx 45°$). These results show that the linear evolution of w with D observed by Deegan et al. for isolated drops deposited manually[16] applies to the case of fused drops deposited by inkjet printing under very different conditions (i.e. nature of particle/substrate, presence or not of co-solvents, particle size/concentration, shape of the deposit) with $D \sim W$ even if not all the particles are going to settle in the external ring. Incidentally, this observation indicates that the thinness of the external lines obtained in this study, compared to past studies, would only be related to the small size of the deposited drops. This small size is probably related to the small nozzle size used in this study (i.e. 20 μm) compared to [10].

The width of the central deposit also decreases but the quantification of this variation is made difficult by the diffuse nature of the edges of the central deposit. The structure of the central deposit at high $\Delta \tilde{x}$ suggests that the central deposit results of a percolation of the small clusters of nanoparticles forming isolated islands in the interline space.

The purpose of this work was to show how the morphology and dimensional characteristics of linear solid deposits obtained by inkjet-printing of a nanoparticle suspension can be modified by considering drying condition where contact line depinning occurs. We show that such mode of evaporation can be achieved by minimizing particle concentration with pure water and a solvent mixture (i.e. water/ethyleneglycol/n-propanol/ethanol). We selected a peculiar particle concentration which, after drying of an isolated drop, forms two concentric rings in which the particles assemble in a continuous and compact monolayer. Our main result is that stable rivulets could be formed by drop overlap leading to the formation of three continuous lines of uniform widths after drying with two thin lines (~1 μm width) defining the outer edges of the deposit and a thick central line. We reveal how the width of the resulting lines is influenced by drop-to-drop spacing in a predictable manner for a large experimental window knowing the drop size.

We believe that our results offer new possibilities toward the realization of nano/microperiodic structures for next generation plasmonic and electronic based set-up by combination of self-assembly and inkjet-printing. Our study also highlights the interest of studying more systematically the mechanisms of non-uniform pattern formation from dried inkjet-printed rivulets of nanoparticle dispersion by taking the example of what has been undertaken over the last 20 years on the subject of individual drop drying.

**ACKNOWLEDGMENTS**

ANR (Agence Nationale de la Recherche) and CGI (Commissariat à l'Investissement d'Avenir) are acknowledged for their financial support through Labex SEAM (Science and Engineering for Advanced Materials and devices), ANR-10-LABX-096 and ANR-18-IDEX-0001". We acknowledge Stephan Suffit (laboratory "Matériaux et Phénomènes Quantiques") for his assistance with SEM-FEG and profilometry in the MPQ clean room. We acknowledge Vincent Humblot ("Franche-Comté Electronique Mécanique Thermique et Optique – Sciences et Technologies" institute) for his assistance for wetting angle measurements.

**SUPPORTING INFORMATION**

Description of the materials and methods associated to Table 1.